\documentclass[journal,transmag]{IEEEtran}

\usepackage[]{graphicx}
\usepackage{amsmath}
\usepackage{url}

\begin{document}
\title{
Curie temperature of Sm$_2$Fe$_{17}$ and Nd$_2$Fe$_{14}$B: a first-principles study 
}

\author{
\IEEEauthorblockN{
Taro Fukazawa\IEEEauthorrefmark{1,3},
Hisazumi Akai\IEEEauthorrefmark{2,3},
Yosuke Harashima\IEEEauthorrefmark{1,3},
Takashi Miyake\IEEEauthorrefmark{1,3}
}
\IEEEauthorblockA{
\IEEEauthorrefmark{1}
CD-FMat,
National Institute
of Advanced Industrial Science and Technology,
Tsukuba, Ibaraki 305-8568, Japan}
\IEEEauthorblockA{
\IEEEauthorrefmark{2}
The Institute for Solid State Physics,
The University of Tokyo,
Kashiwano-ha, Kashiwa,
Chiba 277-8581, Japan}
\IEEEauthorblockA{
\IEEEauthorrefmark{3}
ESICMM, National Institute for Materials Science,
Tsukuba, Ibaraki 305-0047, Japan}
}

\IEEEtitleabstractindextext{%
\begin{abstract}
We calculate intersite magnetic couplings for Sm$_2$Fe$_{17}$,  Nd$_2$Fe$_{14}$ and Nd$_2$Fe$_{14}$X (X = B, C, N, O, F) using Liechtenstein's formula
on the basis of first-principles calculation, and analyze them to investigate the Curie temperature of Sm$_2$Fe$_{17}$ and Nd$_2$Fe$_{14}$B.
We find that the magnetic coupling in the dumbbell bond is strongly ferromagnetic in our calculation, which is against a previous conjecture explaining the low Curie temperature for Sm$_2$Fe$_{17}$. The calculated values of the couplings explain the experimentally observed difference in the Curie temperature of Sm$_2$Fe$_{17}$ and Nd$_2$Fe$_{14}$B.
We also address boron's effects on the Curie temperature of Nd$_2$Fe$_{14}$B,
especially in connection with Kanamori's theory of cobaltization.
\end{abstract}

}

\maketitle
\IEEEdisplaynontitleabstractindextext

\section{Introduction}
  In his development of the Nd--Fe--B magnet, Sagawa had a working hypothesis: if the shortest Fe--Fe bond in R$_2$Fe$_{17}$ (R: rare-earth element) could be elongated, its ferromagnetic state would become more stable.\cite{Sagawa12} Although he used boron with the intention of stretching the bond, the resulting crystal, Nd$_2$Fe$_{14}$B, has an utterly different structure from R$_2$Fe$_{17}$. In the present paper, we address two questions related to this episode---whether the hypothesis is valid and how boron affects the Curie temperature ($T_\mathrm{C}$) in Nd$_2$Fe$_{14}$B.

Later works seemingly support the hypothesis. Li and Morrish deduced an antiferromagnetic exchange for the bond from their experimental results by considering the exchange couplings between neighboring atoms alone and attributed the antiferromagnetic coupling to the shortness of the bond.\cite{Li97} Introduction of Cr is considered as another way to eliminate the antiferromagnetic coupling. Enhancement of $T_\mathrm{C}$ in R$_2$(Fe,Cr)$_{17}$ caused by Cr has been explained along this line.\cite{Hao96,Girt97}

Role of the boron in Nd$_2$Fe$_{14}$B has also attracted attention. Kanamori proposed a concept called cobaltization, in which Fe atoms are made Co-like by boron or another element.\cite{Kanamori90} Tatetsu et al. have recently reported on the basis of first-principles calculation that cobaltization exists in Nd$_2$Fe$_{14}$B and the Co-like Fe elements make the magnetic moments of their neighboring Fe elements larger. However, the net effect from cobaltization is negative to the magnetization because the loss of the moment in the Co-like elements is large. This mechanism must affect the Curie temperature but they let the relation of cobaltization to $T_\mathrm{C}$ remain open.\cite{Tatetsu18}

In this paper, we calculate inter-site magnetic couplings for Sm$_2$Fe$_{17}$ and Nd$_2$Fe$_{14}$X (X = B, C, N, O, F) using Liechtenstein's formula\cite{Liechtenstein87} on the basis of first-principles calculation
to tackle the questions above.
Contrary to the previous results\cite{Li97}, the magnetic coupling between the Fe atoms in the shortest bond of Sm$_2$Fe$_{17}$ is found ferromagnetic in our calculation. The conclusion of Li et al. could be a misattribution of antiferromagnetic couplings of longer bonds.
As for $T_\mathrm{C}$ of Nd$_2$Fe$_{14}$B, we consider metastable Nd$_2$Fe$_{14}$ as a reference. By separating chemical effects from the effect of volume changes, we show that there is a positive contribution to $T_\mathrm{C}$ from the chemical effect of B. We also compare the effect of X on $T_\mathrm{C}$ in Nd$_2$Fe$_{14}$X (X = B, C, N, O, F) to see the chemical trends.

\section{Methods}
We use the Korringa--Kohn--Rostoker Green function method\cite{Korringa47,Kohn54}
for solving the Kohn--Sham equation\cite{Kohn65}
based on density functional theory.\cite{Hohenberg64}
The local density approximation is used in the calculation;
the spin--orbit coupling at the rare-earth site is taken into account with the f-electrons
treated as a trivalent open core for which the configuration is constrained by Hund's rule;
the self-interaction correction\cite{Perdew81} is also applied to the
f-orbitals.
Numbers of sampling points in the full first Brillouin zone are
$12\times 12\times 12$ for Sm$_{2}$Fe$_{17}$ and
$6 \times 6\times 4$ for Nd$_{2}$Fe$_{14}$X.

We assume the Th$_{2}$Zn$_{17}$ structure [space group: R$\bar{3}$m (\#166)] for Sm$_{2}$Fe$_{17}$ and 
the Nd$_2$Fe$_{14}$B structure [space group: P4$_2$/mnm (\#136)] for Nd$_2$Fe$_{14}$
and Nd$_2$Fe$_{14}$X (X = B, C, N, O, F).
We use the lattice parameters from Harashima et al.\cite{Harashima18} for Sm$_{2}$Fe$_{17}$,
and those from Tatetsu et al.\cite{Tatetsu18} for Nd$_2$Fe$_{14}$X.

We use Liechtenstein's method\cite{Liechtenstein87} to estimate intersite magnetic couplings $J_{i,j}$, which is calculated from the energy shifts caused by spin-rotational perturbation on the local potentials at the $i$th and $j$th site.
We also estimate the Curie temperature within the mean-field approximation from the intersite magnetic couplings with considering the classical Heisenberg Hamiltonian:
\begin{equation}
 \mathcal{H}
  =
  -
  \sum_{i,j}
  J_{i,j}
  \vec{e}_i
  \cdot
  \vec{e}_j,
\end{equation}
where $\vec{e}_i$ is a unit vector that denotes the direction of the local spin moment, and the magnitude of the spin moments and the double counting of the bonds are considered in the definition of $J_{i,j}$.
With this definition, the coupling is ferromagnetic when 
$J_{i,j}$ is positive.

\section{Results and Discussion}
\begin{figure}
\centering
 \includegraphics[width=7cm,bb=0 0 504 361]{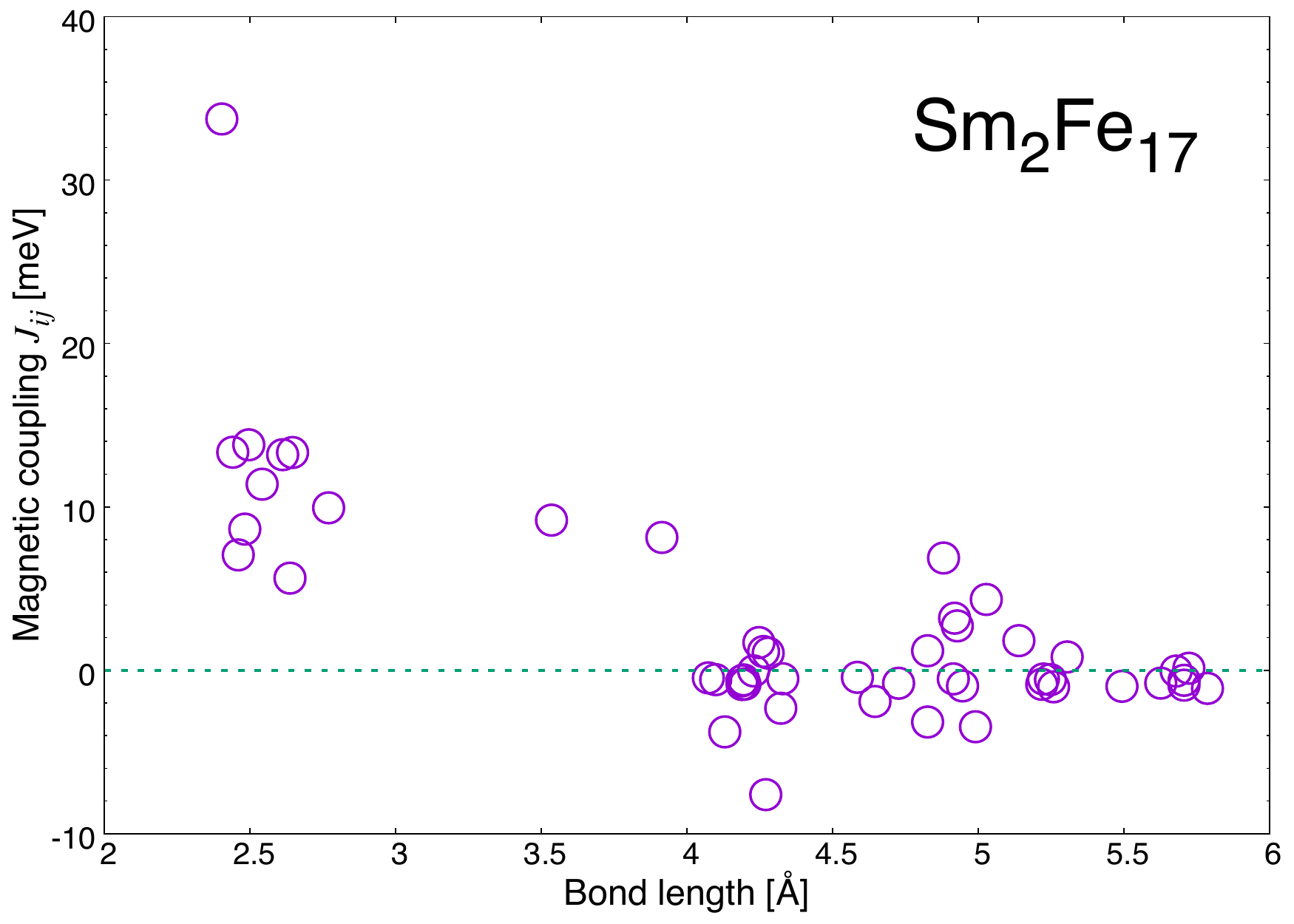}
\caption{
Intersite magnetic couplings $J_{i,j}$
between Fe sites
in Sm$_2$Fe$_{17}$.
The horizontal axis shows the distance between
the $i$th and $j$th site.
\label{Jij_Sm2Fe17}
}
\end{figure}

First, we show the results for Sm$_2$Fe$_{17}$.
Figure~\ref{Jij_Sm2Fe17} shows intersite magnetic couplings
between Fe sites
obtained for Sm$_2$Fe$_{17}$.
Opposing against the previous theory,
the dumbbell bond, which is the shortest one, has a conspicuously
large coupling of approximately 35 meV.
Antiferromagnetic couplings appear only in the region
$R_{i,j} > 4 \text{\ \AA}$ where $R_{i,j}$ denotes the 
bond length.

\begin{figure}
\centering
 \includegraphics[width=8cm,bb=0 0 504 361]{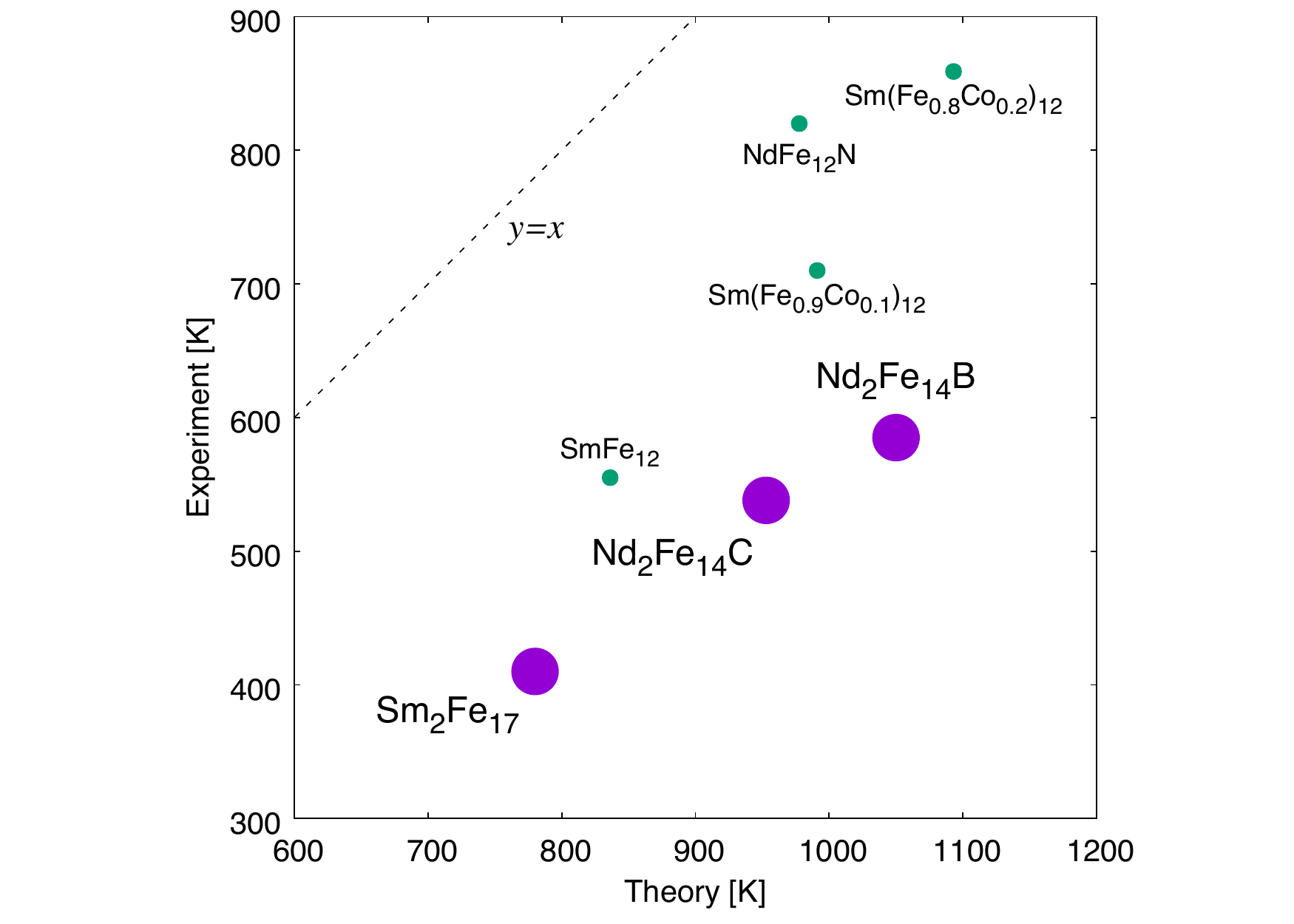}
\caption{
Experimental values of the Curie temperature
versus theoretical values within the mean-field approximation
for Sm$_{2}$Fe$_{17}$,\cite{Li97}
Nd$_2$Fe$_{14}$B, and Nd$_2$Fe$_{14}$C.\cite{Coehoorn89}
The experimental values for Sm(Fe$_{1-x}$Co$_x$)$_{12}$ ($x=0, 0.1, 0.2$),\cite{Hirayama17} and NdFe$_{12}$N\cite{Hirayama15}
are also plotted with theoretical values\cite{Fukazawa17,Fukazawa19}
for comparison.
\label{Tc_comparison}
}
\end{figure}

The estimated Curie temperature ($T_\mathrm{C}$)
is approximately 780 K. Due to the use of the mean-field approximation,
this is overestimated compared to the experimetal value of 
410 K from Li and Morrish.\cite{Li97}
Although the absolute value of $T_\mathrm{C}$ does not agree with 
the experimental one, relative changes of $T_\mathrm{C}$
caused by small alteration of systems
has been adequately reproduced within the mean-field approximation
in previous studies.\cite{Fukazawa18,Fukazawa17,Fukazawa19}
The comparison of theoretical and experimental values of
the Curie teperature
for Sm$_2$Fe$_{17}$, Nd$_2$Fe$_{14}$B and Nd$_2$Fe$_{14}$C
are shown in Fig.~\ref{Tc_comparison},
compared with those for NdFe$_{12}$N and Sm(Fe,Co)$_{12}$ in the
previous studies.

\begin{figure}
\centering
\includegraphics[width=7cm]{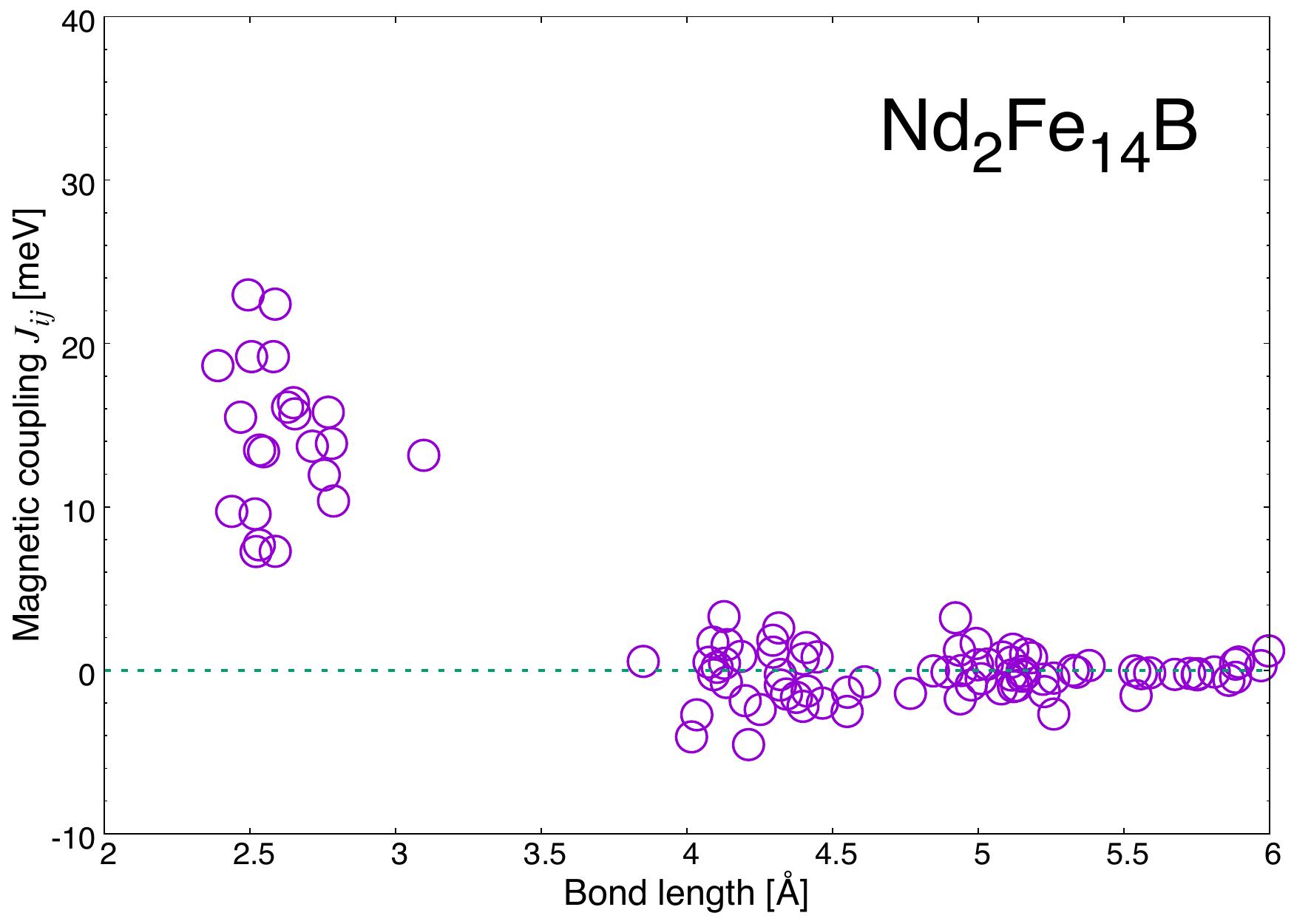}
 \caption{
Intersite magnetic couplings $J_{i,j}$
between Fe sites
in Nd$_2$Fe$_{14}$B.
The horizontal axis shows the distance between
the $i$th and $j$th site.
\label{Jij_Nd2Fe14B}
 }
\end{figure}

Next, we compare those intersite magnetic couplings
to those for Nd$_2$Fe$_{14}$B.
Figure \ref{Jij_Nd2Fe14B} shows
values of the couplings
between Fe sites
obtained for Nd$_2$Fe$_{14}$B.
The overall behavior is similar to
that in Sm$_2$Fe$_{17}$.
Although there are apparently more
data points in Fig. \ref{Jij_Nd2Fe14B} than
Fig. \ref{Jij_Sm2Fe17},
this comes from the difference of the symmetry between
Nd$_2$Fe$_{14}$B and Sm$_2$Fe$_{17}$.

Although each of the two systems has a different
crystal structure from each other,
the geometry affects the Curie temperature
only via the values of the magnetic couplings, $J_{ij}$,
and the definition of sublattices
within the mean-field approximation.
To see how the sublattices affects the Curie temperature,
let us consider
a further approximation to the mean-field 
approach by regarding all the Fe sites as 
of a single sublattice and disregarding all the other
elements.
The Curie temperature within the mean-field approximation
is obtained as 
the largest eigenvalue of the linear equation\cite{Fukazawa18}
for the Weiss field $\vec{m}_i$:
\begin{equation}
 \vec{m}_i = \sum_j \frac{2}{3k_\mathrm{B}} J_{i,j} \vec{m}_j.
 \label{Weiss}
\end{equation}
Because the local magnetic moments of the Fe sites
in those systems
are coupled strongly and ferromagnetically,
the Weiss fields at a temperature 
slightly below the Curie temperature
can be efficiently approximated by 
$\vec{m}_i \simeq \delta \vec{e}_z$
where $\delta$ is a small number that does not
depend on $i$. This leads to a rough estimation of
the mean-field Curie temperature:
$T_\mathrm{C}^\text{rough-MFA} =  2/(3k_\mathrm{B}N) \sum_{i,j} J_{ij}$,
the values of which are
770 K for Sm$_2$Fe$_{17}$,
990 K for Nd$_2$Fe$_{14}$B, and
890 K for Nd$_2$Fe$_{14}$C
while they are 780 K, 1050 K, 950 K
within the oridinary mean-field approximation, repsectively.
Because the effects of the sublattices are minor,
we can directly compare $J_{ij}$ shown
in Fig.~\ref{Jij_Sm2Fe17} and Fig.~\ref{Jij_Nd2Fe14B}
to discuss their Curie temperatures.

Most part of the Curie temperature can be explained
by the bonds with the lengths smaller than 3~\AA.
Value of $T_\mathrm{C}$ for Nd$_2$Fe$_{14}$B
with the cutoff of 3~\AA\ is 1100 K while
it is 1050 K when the all the bonds less than 7 \AA\ 
are considered.
In the case of Sm$_2$Fe$_{17}$,
$T_\mathrm{C}$ with the cut off of 
3~\AA\ is 870 K while it is 780 K with
all the bonds less than 10 \AA.

On this grounds, we consider
substantial difference is
the existence of
the magnetic couplings in the range of 15--25 meV
with the bond lengths smaller than 3~\AA,
which can be classified
by the Wycoff positions of Nd$_2$Fe$_{14}$B
[P4$_2$/mnm (\#136)] as in 
Table \ref{table_Nd2Fe14B_couplings}.
\begin{table}[htbp]
 \caption{Intersite magnetic couplings in
 {\rm Nd$_2$Fe$_{14}$B}
 in the range of 15--25 {\rm meV}
}
\label{table_Nd2Fe14B_couplings}
\centering
\begin{tabular}{lc|lc}
 \hline
 \hline
 Bond  & J (meV) & Bond& J (meV)\\
 \hline
 Fe(16k$_1$)--Fe(16k$_1$) & 22.4 & Fe(16k$_2$)--Fe(8j$_2$) & 16.1\\
 Fe(16k$_1$)--Fe(16k$_2$) & 15.5 & Fe(16k$_2$)--Fe(8j$_2$)*& 15.7\\
 Fe(16k$_1$)--Fe(4e)      & 23.0 & Fe(16k$_2$)--Fe(4c)     & 19.2 \\ 
 Fe(16k$_1$)--Fe(4c)      & 19.2 & Fe(8j$_1$)--Fe(8j$_2$)  & 16.4\\
 Fe(16k$_2$)--Fe(8j$_1$)  & 18.6 & Fe(8j$_1$)--Fe(8j$_2$)* & 15.8\\
 \hline
 \hline
 \multicolumn{4}{l}{The asterisks denote the second shortest bonds in
 the classes.} \\
 \multicolumn{4}{l}{All the others refer to the shortest.}
\end{tabular}
 \end{table}
Although the ferromagnetic coupling of the dumbbell bond
in Sm$_2$Fe$_{17}$ is strong,
there are only two pairs in the unit cell
and they cannot compensate for the difference.

\begin{figure}
\centering
\includegraphics[width=7cm,bb=050 050 791 544]{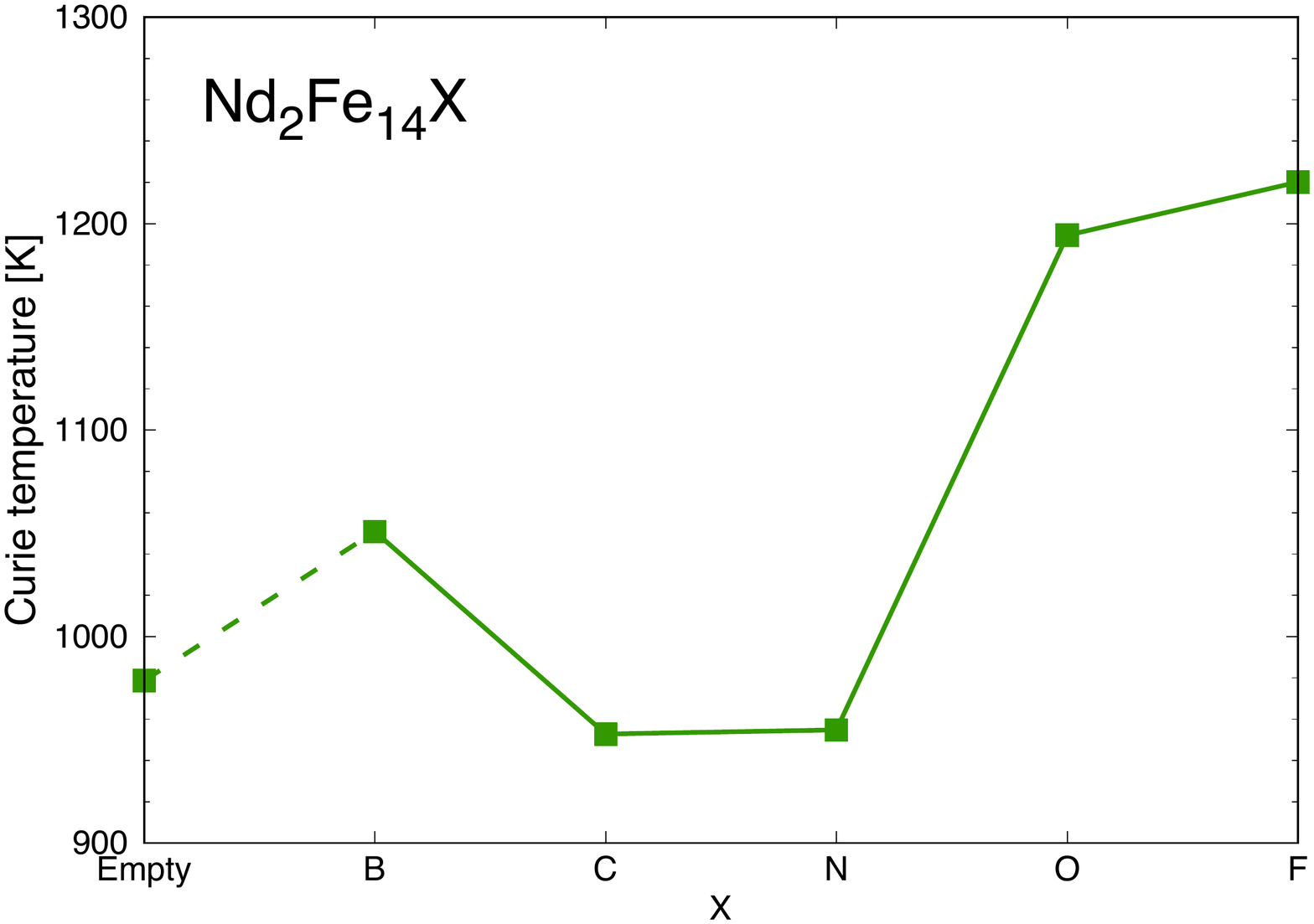}
 \caption{
 Values of the Curie temperature within the mean-field approximation
 for Nd$_{2}$Fe$_{14}$ (shown as X = Empty) and
 Nd$_{2}$Fe$_{14}$X (X = B, C, N, O, F).
 \label{Tc_Nd2Fe14X}
 }
\end{figure}

Finally, we show our results of the Curie temperature
for Nd$_2$Fe$_{14}$ and
Nd$_2$Fe$_{14}$X (X = B, C, N, O, F) within
the mean-field approximation.
Figure~\ref{Tc_Nd2Fe14X} shows
values of $T_\mathrm{C}$ for them.
It is noteworthy that the estimated value of $T_\mathrm{C}$
for Nd$_2$Fe$_{14}$ is approximately 980 K and is higher than that of
Sm$_2$Fe$_{17}$ by $~$200 K
while the difference of $T_\mathrm{C}$ for Nd$_2$Fe$_{14}$B
from Sm$_2$Fe$_{17}$ is $~$270 K.
Therefore, large part of difference of $T_\mathrm{C}$
can be attributed to the difference in the structure
between the compounds.

As shown in the figure, there still is
a change of a few hundred Kelvin in
$T_\mathrm{C}$ as a function of
the element X.
Kanamori pointed out the importance of hybridization between
the d-states of Fe elements and states of the neighboring typical
elements in understanding magnetism of compounds.\cite{Kanamori90}
He also suggested the possibility of the hybridization for enhancing
the ferromagnetism in Nd$_2$Fe$_{14}$B.
It is called cobaltization because the typical elements
make the neighboring Fe elements Co-like by the hybridization.

Tatetsu et al. \cite{Tatetsu18}
in their analysis
decomposed 
the effect of typical element X
in Nd$_2$Fe$_{14}$X (X = B, C, N, O, F)
into a magnetovolume effect $\Delta V$
and a chemical effect $\Delta C$
by considering the metastable
Nd$_2$Fe$_{14}$
and a non-stable system of Nd$_2$Fe$_{14}$
having the lattice parameter of Nd$_2$Fe$_{14}$X.
Let us denote the latter system by
Nd$_2$Fe$_{14}${\scriptsize \#Nd$_2$Fe$_{14}$X}.
The magnetovolume effect, $\Delta V$, is
defined as the
difference of the property of 
Nd$_2$Fe$_{14}${\scriptsize \#Nd$_2$Fe$_{14}$X} from that of
Nd$_2$Fe$_{14}$;
the chemical effect, $\Delta C$, is defined as the
difference of the property of
Nd$_2$Fe$_{14}$X
from that of Nd$_2$Fe$_{14}${\scriptsize \#Nd$_2$Fe$_{14}$X},
which is summarized in the following
diagram:
\begin{equation}
 \begin{array}{ccc}
  \text{Nd$_2$Fe$_{14}$X{\scriptsize \#Nd$_2$Fe$_{14}$}}& \overset{\Delta V +
   \alpha}{\longrightarrow} & \text{Nd$_2$Fe$_{14}$X}\\
  \\
  {\scriptstyle \Delta C-\alpha}\Big\uparrow& & \Big\uparrow
   {\scriptstyle \Delta C} \\
  \\
  \text{Nd$_2$Fe$_{14}$} & \underset{\Delta V}{\longrightarrow}& \text{Nd$_2$Fe$_{14}${\scriptsize \#Nd$_2$Fe$_{14}$X}}\\
 \end{array}
\end{equation}
It was found in the previous study
that $\Delta C$ of the magnetization
in units of tesla
was negative for X = B.
On this account, it was concluded that
the cobaltization
worked to decrease the magnetization in total,
although some local moments were increased by the effect.

\begin{figure}
\centering
\includegraphics[width=7cm]{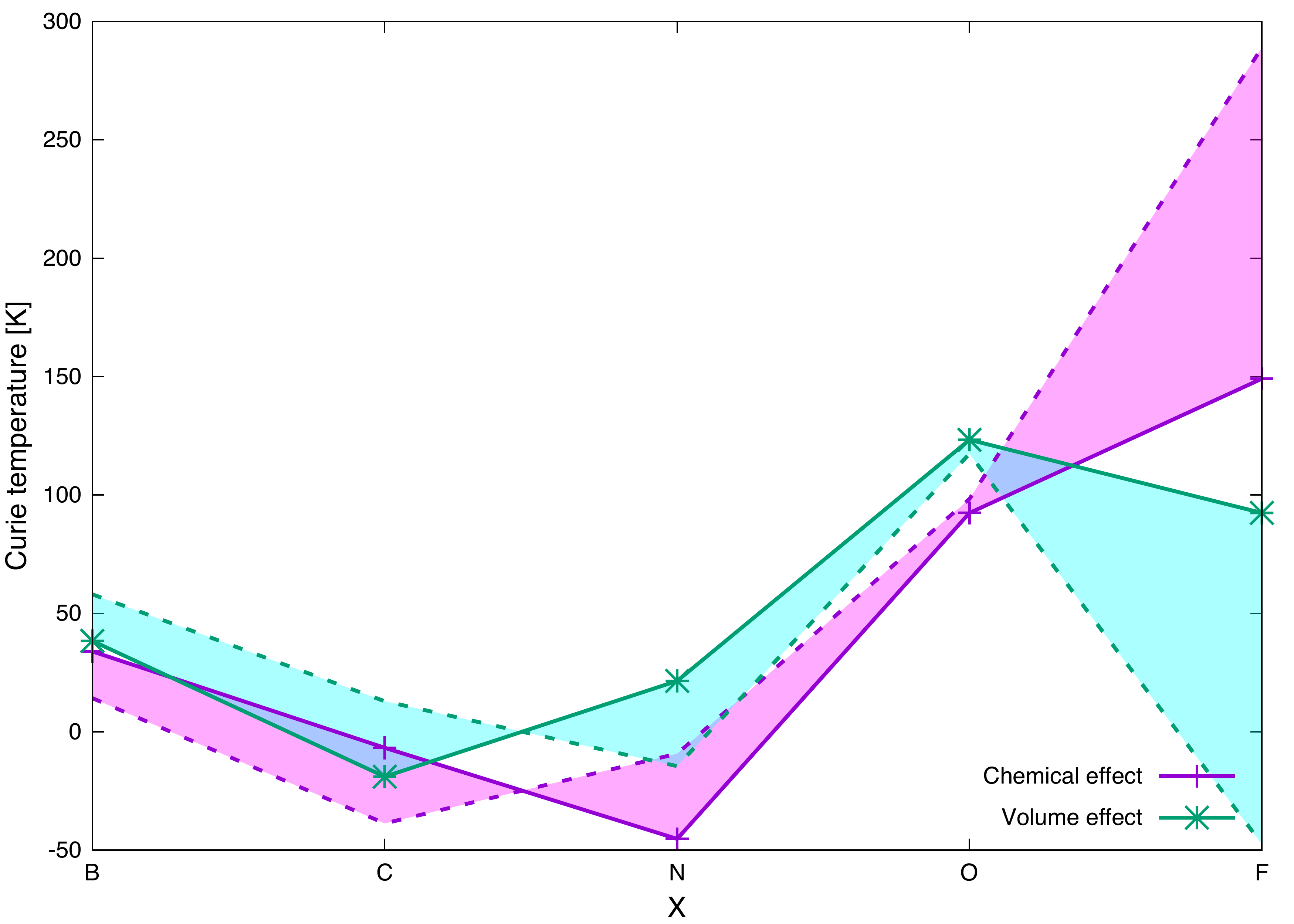}
 \caption{
 Values of the chemical effect, $\Delta C$,
 and the magnetovolume effect, $\Delta V$, for
 the Curie temperature within the mean-field approximation
 for
 Nd$_{2}$Fe$_{14}$X (X = B, C, N, O, F).
 The dotted lines show
 the values of $\Delta C - \alpha$ and
 $\Delta V + \alpha$.
 The widths of the color bands show the magnitude of
 $\alpha$.
 \label{cvalpha}
 }
\end{figure}

We here apply this decomposition to $T_\mathrm{C}$
to see the role of X in changing the Curie temperature.
Figure \ref{cvalpha} shows values of
$\Delta V$, $\Delta C$ and $\alpha$
for the Curie temperature of Nd$_2$Fe$_{14}$X
as functions of X.
In Nd$_2$Fe$_{14}$B,
the chemical effect is positive.
Therefore, the hybridization between Fe and B is considered to work
to enhance
the Curie temperature by a few tens of Kelvin.
We can see from Fig.~\ref{cvalpha} that
there are considerable chemical effects
for X = O and X = F.
This corresponds to occupation of
the antibonding states between
Fe and X sites, which was previously
pointed out by Tatetsu et al
to explain the magnetization.\cite{Tatetsu18}
The change in the Curie temperature
can be attributed to
this change in the electronic structure also
for X = O and F.

The definition of the chemical and magnetovolume
effect above is adequate because
$\Delta V$ is a change of the property when
the structure is changed with fixing
the chemical formula and
$\Delta C$ is a change when the element X is
added without changing the structure.
However, at the same time,
there is another possible definition
that satisfies the statements above.
Let us denote
a non-stable system of Nd$_2$Fe$_{14}$X
having the lattice parameters of
Nd$_2$Fe$_{14}$
by Nd$_2$Fe$_{14}$X{\scriptsize \#Nd$_2$Fe$_{14}$}.
Then, $\Delta V + \alpha$ and $\Delta C - \alpha$
in the diagram above
also can serve as estimations of 
the magnetovolume and chemical effect, respectively.
(Note that the difference of the property between
Nd$_2$Fe$_{14}$X and Nd$_2$Fe$_{14}$ is
$\Delta V + \Delta C$ anyway).
We can regard this $\alpha$ as an estimation of the ambiguity
in $\Delta V$ and $\Delta C$.
Another possible interpretation is that $\alpha$ is
a combined effect of the structural and chemical change:
the structural change from Nd$_2$Fe$_{14}$ yields
$\Delta V$; the chemical change from Nd$_2$Fe$_{14}$
yields $\Delta C - \alpha$;
the extra change $\alpha$ emerges when those two operations
are combined.

The value of $\alpha$ is shown in Fig.~\ref{cvalpha}.
The chemical effect in Nd$_2$Fe$_{14}$B is smaller
in the alternative definition but is still positive.
With the exception of Nd$_2$Fe$_{14}$F, the absolute values
of $\alpha$ are approximately 50 K or less.
That for X = F is extremely large 
due to the large difference in structure from Nd$_2$Fe$_{14}$.

\section{Conclusion}
We discussed intersite magnetic couplings
in Sm$_2$Fe$_{17}$ and Nd$_2$Fe$_{14}$B on the basis
of first-principles calculation.
In contrast to the previous theory,
we found the dumbbell bond (the smallest Fe--Fe bond) in
Sm$_2$Fe$_{17}$ strongly ferromagnetic.
We attribute 
the smallness of the Curie temperature in Sm$_2$Fe$_{17}$
to lack of ferromagnetic couplings in the range of 15--25 meV.
We also discuss the role of boron in enhancing the Curie temperature.
We could say that the largest contribution of boron is
stabilizing the Nd$_2$Fe$_{14}$B structure, with which
$T_\mathrm{C}$ becomes larger than Sm$_2$Fe$_{17}$
by $~$200 K even without boron (Nd$_2$Fe$_{14}$).
We also analyzed how cobaltization can enhance the Curie temperature.
We found that there is a positive contribution of a few tens of Kelvin
from the chemical effect of boron.
The difference in the electronic strcuture
that was observed for X = O and F, contributes to
the Curie temperature.

\section*{Acknowledgment}

The authors would like to thank Yasutomi Tatetsu, Yuta Toga and Shotaro Doi for sound advice
and fruitful discussions.
We gratefully acknowledge the support from the Elements Strategy Initiative Project under the auspices of MEXT. This work was also supported by MEXT as a social and scientific priority issue (Creation of new functional Devices and high-performance Materials to Support next-generation Industries; CDMSI) to be tackled by using the post-K computer. 
The computation was partly conducted using the facilities of the Supercomputer Center, the Institute for Solid State Physics, the University of Tokyo, and the supercomputer of ACCMS, Kyoto University. 
This research also used computational resources of the K computer provided by the RIKEN Advanced Institute for Computational Science through the HPCI System Research project (Project ID:hp170100).

\appendices
\section{Local moments of {\rm Sm$_2$Fe$_{17}$}
and {\rm Nd$_2$Fe$_{14}$X}}

Table \ref{local_moments_Sm2Fe17} shows 
calculated local moments of Sm$_2$Fe$_{17}$,
where the crystal symmetry of 
R$\bar{3}$m (\#166) is assumed.
\begin{table}[htbp]
\renewcommand{\arraystretch}{1.3}
\caption{Local magnetic moments of {\rm Sm$_2$Fe$_{17}$}
in units of $\mu_\mathrm{B}$
}
 \label{local_moments_Sm2Fe17}
\centering
\begin{tabular}{rccccc}
 \hline
 \hline
                  & Sm & Fe(6c) & Fe(9d) & Fe(18f) & Fe(18h)  \\
 \hline
  Sm$_2$Fe$_{17}$ & -0.362 & 2.52 & 1.81 & 2.33 & 2.25  \\
 \hline
\hline
\end{tabular}
\end{table}

Table \ref{local_moments_Nd2Fe14X} shows 
calculated local moments of Nd$_2$Fe$_{14}$ and
Nd$_2$Fe$_{14}$X (X = B, C, N, O, F),
which is assumed to have the symmetry of 
P4$_2$/mnm (\#136).
\begin{table}[htbp]
\renewcommand{\arraystretch}{1.3}
\caption{Local magnetic moments of
{\rm Nd$_2$Fe$_{14}$X}
in units of $\mu_\mathrm{B}$
}
 \label{local_moments_Nd2Fe14X}
\centering
\begin{tabular}{rccccccccc}
 \hline
 \hline
                  & Nd(4f) & Nd(4g) & Fe(4e) & Fe(4c) & X  \\
 \hline
  Nd$_2$Fe$_{14}$B & -0.422 & -0.263 & 1.80 & 2.50 & -0.164 \\
  Nd$_2$Fe$_{14}$C & -0.696 & -0.588 & 1.58 & 2.24 & -0.186 \\
  Nd$_2$Fe$_{14}$N & -0.395 & -0.230 & 1.89 & 2.16 & -0.006 \\
  Nd$_2$Fe$_{14}$O & -1.088 & -0.203 & 2.60 & 2.39 &  0.154 \\
  Nd$_2$Fe$_{14}$F & -0.366 & -0.133 & 2.57 & 2.56 &  0.145 \\
  Nd$_2$Fe$_{14}$  & -0.430 & -0.400 & 2.50 & 2.02 & -0.009 \\
 \hline
\hline
\vspace{0.0625in}\\
 \hline
 \hline
                  & Fe(16k$_1$) &  Fe(16k$_2$) & Fe(8j$_1$) & Fe(8j$_2$)  & ---\\
 \hline
  Nd$_2$Fe$_{14}$B & 2.07 & 2.19 & 2.34 & 2.65 & ---\\
  Nd$_2$Fe$_{14}$C & 1.97 & 2.14 & 2.27 & 2.65 & --- \\
  Nd$_2$Fe$_{14}$N & 2.23 & 2.00 & 2.24 & 2.59 & --- \\
  Nd$_2$Fe$_{14}$O & 2.55 & 1.97 & 2.32 & 2.52 & --- \\
  Nd$_2$Fe$_{14}$F & 2.55 & 2.00 & 2.30 & 2.46 & --- \\
  Nd$_2$Fe$_{14}$  & 2.43 & 2.06 & 2.14 & 2.54 & --- \\
 \hline
\hline
\end{tabular}
\end{table}

\section{Curie temperature of {\rm Sm$_2$Fe$_{17}$}
with its dumbbell bond elongated}
\begin{figure}
\centering
\includegraphics[width=7cm,bb=000 000 504 360]{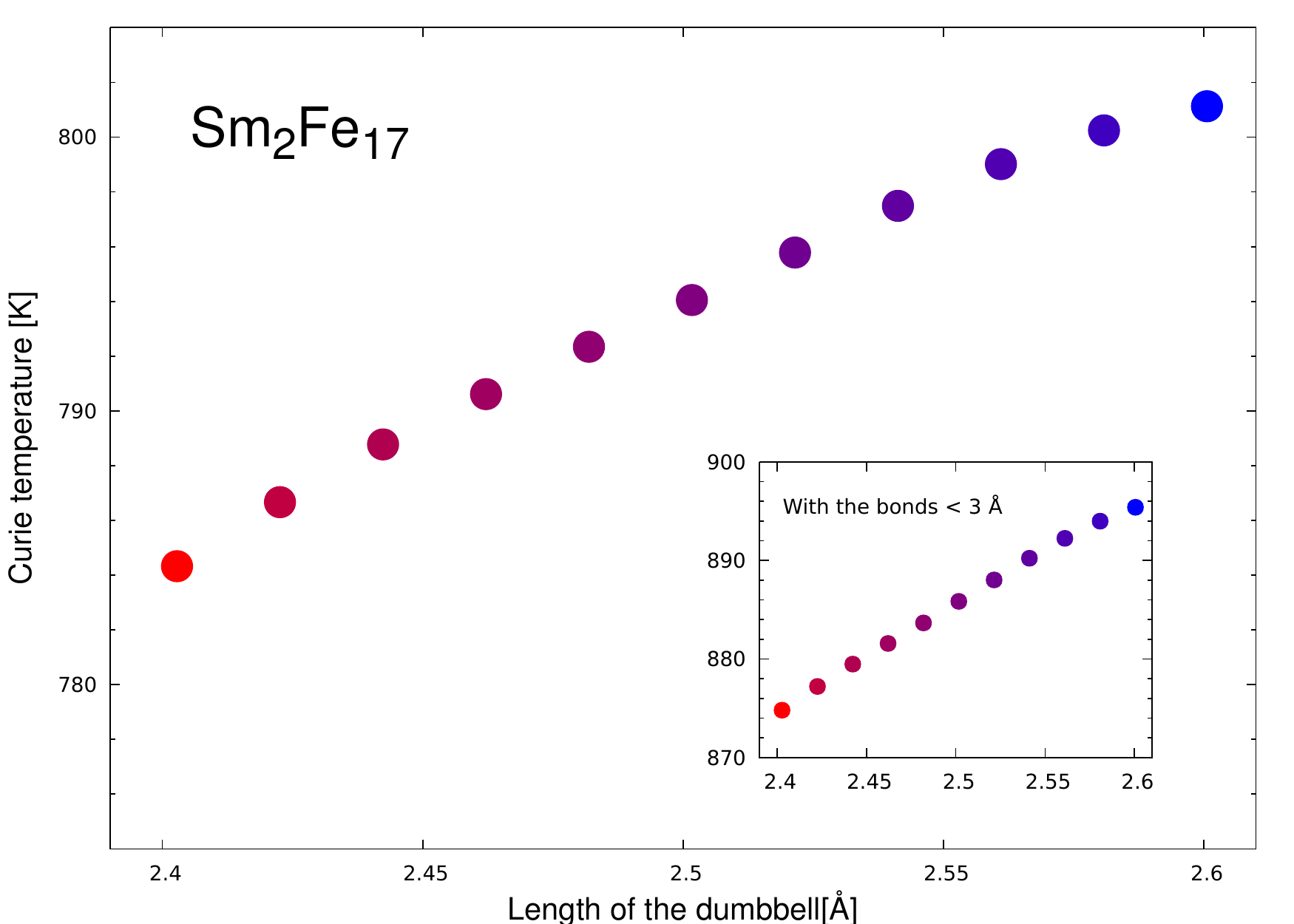}
 \caption{
 The Curie temperature of Sm$_2$Fe$_{17}$
 within the mean-field
 approximation when
 the dumbbell bond is elongated by changing 
 an inner parameter keeping the Th$_2$Zn$_{17}$ structure.
 Those obtained from the bonds less than 3 \AA\ are also 
 shown in the inset.
 The colors of the data points correspond to 
 those in Fig.~\ref{Jij_Sm2Fe17_elongated}.
 \label{Tc_Sm2Fe17_elongated}
 }
\end{figure}

Figure \ref{Tc_Sm2Fe17_elongated} shows
the Curie temperature of Sm$_2$Fe$_{17}$
with its dumbell bond elongated.
In these calculations, we increased an inner parameter
of the lattice to elongate the dumbbell bond 
with the symmetry of the Th$_2$Zn$_{17}$ structure
kept.
The leftmost red point in the figure corresponds to 
the data for Sm$_2$Fe$_{17}$ shown in the main body.

\begin{figure}
\centering
\includegraphics[width=7cm,bb=000 000 504 360]{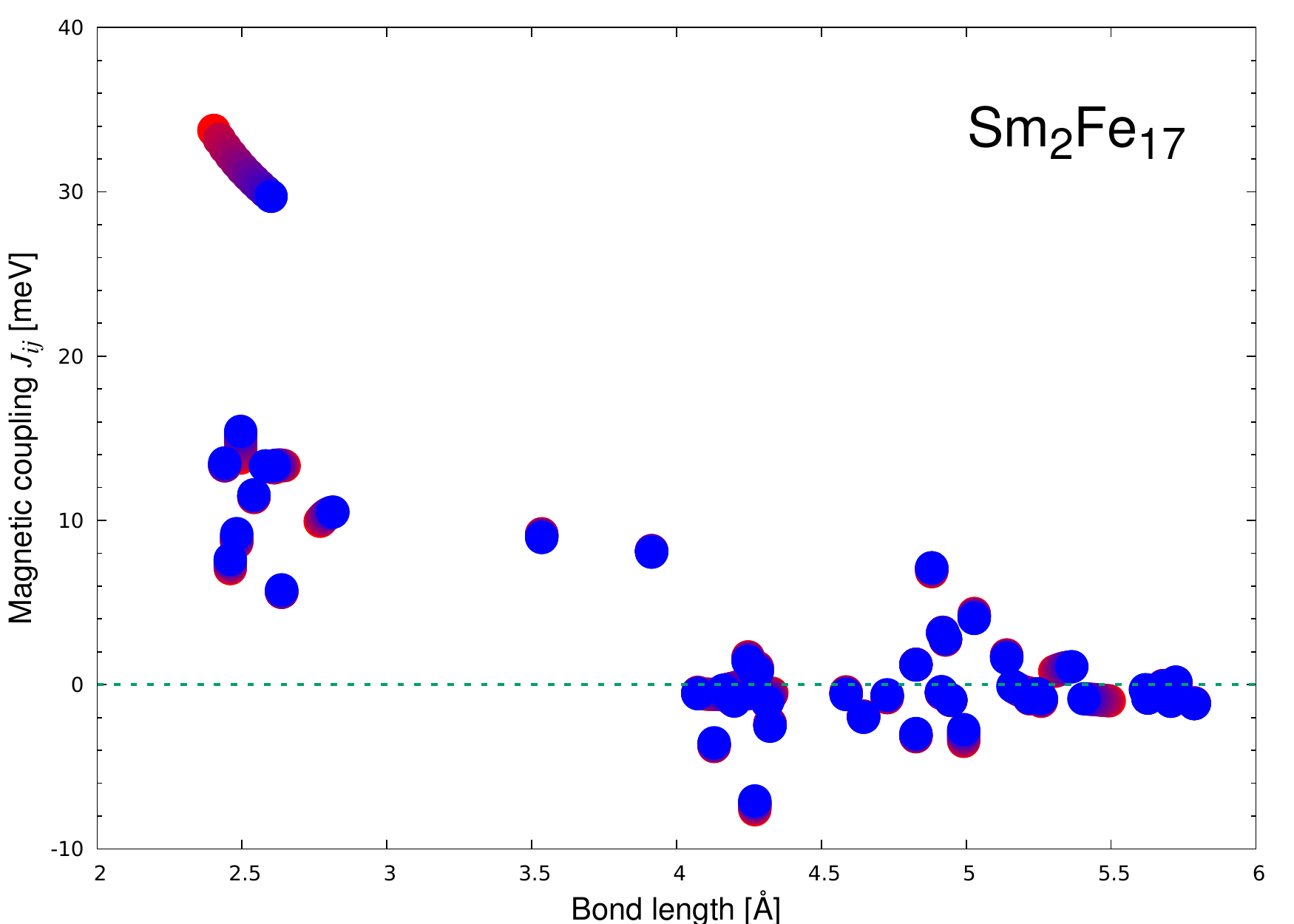}
 \caption{Intersite magnetic couplings $J_{i,j}$ between Fe sites
 in Sm$_2$Fe$_{17}$ when the dumbbell bond is elongated by changing 
 an inner parameter, keeping the Th$_2$Zn$_{17}$ structure.
 The horizontal axis shows the distance between
 the $i$th and $j$th site.
 The red and blue colors correspond to $J_{i,j}$'s for different length
 of the dumbbell bond.
 \label{Jij_Sm2Fe17_elongated}
 }
\end{figure}

Figure \ref{Jij_Sm2Fe17_elongated} shows
the intersite magnetic couplings in 
these calculations.
It is noteworthy that 
the magnetic coupling between the dumbbell bond
decreases as it is elongated
while the Curie temperature increases.
Because only two dumbell bonds exist per unit cell,
the change in the magnetic coupling does not affect the 
Curie temperature much.
The values calculated from 
the bonds less than 
3 \AA\ are also shown in the inset.
This cutoff does not change
the overall behavior,
which suggests the importance of
the other bonds less than 3 \AA.

\bibliography{main}
\bibliographystyle{IEEEtran}

\end{document}